\documentclass[12pt, final]{l4dc2022}

\title[Neural IDA-PBC]{Total Energy Shaping with Neural Interconnection and Damping Assignment - Passivity Based Control}\usepackage{times}

\author{%
 \Name{Santiago {Sánchez-Escalonilla}} \Email{santiago.sanchez@rug.nl}\\[2pt]
 \Name{Rodolfo Reyes-Báez} \Email{r.reyes.baez@rug.nl}\\[2pt]
 \Name{Bayu Jayawardhana} \Email{b.jayawardhana@rug.nl}\\[2pt]
 \addr Jan C. Willems Center for Systems and Control,\\ Engineering and Technology Institute Groningen (ENTEG),\\ Rijksuniversiteit Groningen%
}

\usepackage{dsfont}
\usepackage{xcolor}
\usepackage{verbatim}
\usepackage{enumerate}
\usepackage{caption}
\usepackage{natbib}
\usepackage{amsmath,amssymb,amsfonts}
\usepackage{amsmath,mathtools}
\usepackage{algorithmic}
\usepackage{graphicx}
\usepackage{textcomp}
\usepackage{xcolor}
\usepackage{natbib}   
\usepackage{amsmath,bm}

\newcommand{\sbm}[1]{\left[\begin{smallmatrix} #1
   \end{smallmatrix}\right]}

\begin{document}

\maketitle

\begin{abstract}%
In this work we exploit the universal approximation property of Neural Networks (NNs) to design interconnection and damping assignment (IDA) passivity-based control (PBC) schemes for fully-actuated mechanical systems in the port-Hamiltonian (pH) framework. To that end, we transform the IDA-PBC method into a supervised learning problem that solves the partial differential matching equations, and fulfills equilibrium assignment and Lyapunov stability conditions. A main consequence of this, is that the output of the learning algorithm has a clear control-theoretic interpretation in terms of passivity and Lyapunov stability. The proposed control design methodology is validated for  mechanical systems of one and two degrees-of-freedom via numerical simulations.
\end{abstract}


\section{Introduction}
In many nonlinear control techniques the construction of the nonlinear controller is conditioned to the explicit solution of a set of partial differential equations (PDEs) that embed the design requirements, for instance,  feedback linearization \citep{isidori1995nonlinear}, Lyapunov-based control \citep{khalil_nonlinear_2015}, and passivity-based control (PBC) \citep{arjanl2}. The latter approach is a popular technique in practice due to the clear physical interpretation of the control scheme in terms of energy. In this context, the  IDA-PBC technique, introduced in  \cite{ortega_interconnection_2002}, is a well-known  PBC design method for the control of affine  nonlinear systems, for which the target closed-loop dynamics take the form of a passive pH system with desired passivity and interconnection properties. A major disadvantage of this method is that to construct the control scheme, it is necessary to solve a set of PDEs, the so-called \emph{matching equations} (MEs), in order to obtain the target pH closed-loop system. While solving these PDEs is non-trivial in general, analytic solutions can be obtained for some classes of systems by restricting the family of target pH systems. We refer interested readers to  
\citep{nageshrao_adaptive_2015} for a review on methods to solve the IDA-PBC PDEs.

In the present paper we propose a method to overcome the problem of finding analytical solutions of the MEs by transforming the IDA-PBC design into a supervised learning problem  using NNs where the minimized cost function is systematically constructed to represent the IDA-PBC methodology and port-Hamiltonian  systems' first-principles, resulting in a control-informed NN hereupon referred as Neural IDA-PBC. This is motivated by  the universal approximation property of NNs \citep{hornik_multilayer_1989,cybenko_approximation_1989,kratsios_universal_2021}. Generally, the main focus of systems \& control theory is the analysis  of  dynamical systems and the design of control input functions, that can steer the system to a desired state satisfying a set of dynamical properties. Correspondingly, it is natural to use NNs as functional surrogates of the control input functions. 
Yet, for the past decades, the application of NNs in the control (such as in data-driven control) of complex engineering systems is fairly limited due to the lack of explainability of the solution, and their use is oftentimes discouraged. Fortunately, recent progress in the regularization of NNs has allowed us to obtain parametric functions that satisfy predefined restrictions. With the seminal work physics-informed neural networks (PINNs) \citep{raissi_physics_2017,raissi_physics_2017-1,karniadakis_physics-informed_2021}, it has been shown that constitutive physical laws (in the form of PDEs) can be encoded in the training process, which results in NNs that “understand/respect” the physics of the problem. In other words, during the optimization process, an approximated solution of the underlying physical laws can be obtained. 

In the past, researchers have studied different ways of combining learning with control theory. As an example, finding the right Control-Lyapunov function with desirable closed-loop properties (such as large basin of attraction) can be cumbersome and often relies in polynomial approaches that limit the control law to a particular function class. \cite{richards_lyapunov_2018,chang_neural_2020, grune_computing_2020} propose using NNs to learn new families of Lyapunov candidates. \cite{chang_stabilizing_2021} show how the Lyapunov method can be used to regularize the policy optimization process in Reinforcement Learning, by adding a Lyapunov Critic into the optimization loop. Although the IDA-PBC method is rewritten as an optimization problem, in the present paper, we do not consider any performance metrics over the trajectory. For a review in the connection between optimal control and PBC, the reader is referred to \citep{fujimoto_optimal_2003, vu_connection_2018, massaroli_optimal_2021}.

The rest of this paper is divided into 4 sections: Section \ref{sec:Preliminaries} briefly introduces mechanical systems in the port-Hamiltonian formulation, followed by the IDA-PBC control methodology as presented in \cite{ortega_interconnection_2002}; Section \ref{sec:NeuralIDA} contains the main contribution of this paper, starting from the universal approximation property of NNs, we define a series of residual terms that encode the fundamentals of IDA and PBC; Section \ref{sec:Simulations} numerically validates the Neural IDA-PBC approach; and finally, Section \ref{sec:Conclusions} closes with conclusions and future work.

\section{Preliminaries}\label{sec:Preliminaries}

\subsection{Mechanical systems in the port-Hamiltonian framework}\label{sec:Preliminaries-pH}
The dynamics of a standard mechanical system  in the pH framework (see   \cite{sanz-sole_port-hamiltonian_2007}) with generalized coordinates $q$ on the  configuration space $\mathcal{Q}\subset\mathds{R}^n$ and velocity $\dot{{q}}\in T_{q}\mathcal{Q}$, is  
	\begin{equation}
	\dot{x}=[J(x)-R(x)]\frac{\partial H}{\partial x}(x)+g(x)u, \qquad
	y=g^{\top}(x)\frac{\partial H}{\partial x}(x),
	\label{eq:phmechanical}
	\end{equation}
with  Hamiltonian function given by the total energy of the system $H({x})=\frac{1}{2}{p}^{\top}{M}^{-1}({q}){p}+U({q})$, where ${x}=({q},{p})\in\mathcal{X}$ is the   state  and  $p:=M(q)\dot{q}$ is the generalized momentum. The scalar function $U({q})$ is the potential energy,   and    $M(q)=M^{\top}(q)>0$ is the   inertia matrix. The interconnection, dissipation and input matrices, respectively are given by 
$	J(x)= \sbm{{0}_n & {I}_n\\
	-{I}_n& {0}_n}$, 
$ R(x)=	\sbm{{0}_n & {0}_n\\
	{0}_n& {D}(x)}$, 
$g(x)= \sbm{{0}_n\\
	B(q)}$, 
where the  $n\times n$ matrix ${D}(x)={D}^{\top}(x) \geq {0}_n$ is a dissipation term; and  $I_n$ and $0_n$ are the $n\times n$ identity and zero matrices, respectively. The input $u$ represents generalized forces while the output $y$ gives the generalized velocity so that their inner-product is power. The input force matrix $B(q)$ has rank $m\leq n$. If $n=m$ then the mechanical system in  \eqref{eq:phmechanical} is  \emph{fully-actuated}.
	
The  rate of change of  $H(x)$ provides a power balance relation between the internal power of system   \eqref{eq:phmechanical} and the external supplied power, the so-called power balance analysis:
\begin{equation}
\dot{H}(x)=\frac{\partial H^{\top}}{\partial q}(x)\dot{q}+\frac{\partial H^{\top}}{\partial p}(x)\dot{p}=-\frac{\partial H^{\top}}{\partial p}(x)D(x)\frac{\partial H}{\partial p}(x)+y^{\top}u\leq y^{\top}u.
\label{eq:Hamltonian-energybalance-chapter}
\end{equation}
where the inner-product $y^\top u$ is the external supplied power. 
In the context of \emph{dissipativity  theory}, the inequality in \eqref{eq:Hamltonian-energybalance-chapter} shows that  the map $u\mapsto y$ is  \emph{passive} with respect to the     storage function given by the Hamiltonian function $H$ and the supply rate $y^{\top}u$. The interested reader is referred to  \cite[Section 6.2]{arjanl2} for a detailed treatment on passivity. 

\subsection{Interconnection and Damping Assignment Passivity-based Control (IDA-PBC)}\label{sec:IDA}
The IDA-PBC nonlinear technique, as proposed in \citep{ortega_interconnection_2002}, is a passivity-based control design method whose main control objective is to design a \emph{static} state feedback control law of the form $u(x) = \beta(x) + v$ for  the pH system in 
\eqref{eq:phmechanical}, such that the closed-loop dynamics is given by the target  passive pH system  
\begin{equation}
		\dot{x}=[J_d(x)-R_d(x)]\frac{\partial H_d}{\partial x}(x) + g(x)v,\qquad 
		y' = g(x)^\top \frac{\partial H_d}{\partial x}(x),
	\label{eqn:system2}
\end{equation}
where   $J_d(x)=-J_d^{\top}(x)$ and $R_d(x)=R_d^{\top}(x)\geq 0$ are  the \emph{desired} interconnection  and  damping matrices, respectively; and $H_d(x)$ is the desired energy function. The desired Hamiltonian function $H_d(x)$ has a  strict local minimum at the desired equilibrium point $x^\star = (q^\star, 0)$ and the tuple $(v,y')$ is the new power conjugate pair that defines the desired passivity relation  with storage function $H_d(x)$. 

The main result in   \cite{ortega_interconnection_2002} is the following:
\begin{theorem}[IDA-PBC]\label{theorem:IDA-PBC}
Consider $J(x), R(x), H(x), g(x)$ of the pH system in \eqref{eq:phmechanical} and the desired equilibrium to be stabilized $x^*\in\mathcal{X}$. Assume that there exists functions $ J_d(x), R_d(x)$ and $H_d(x)$ satisfying   the \emph{matching equation}
\begin{align}
	g^\perp(x)[J_d(x)-R_d(x)]\frac{\partial H_d}{\partial x}(x) = g^\perp(x)[J(x)-R(x)]\frac{\partial H}{\partial x}(x),	\label{eqn:matchingeqs}
\end{align}             
where $g^\perp(x)$ is the full rank left annihilator of $g(x)$. Then the control law 
\begin{equation}
    u= \underbrace{[g^\top(x)g(x)]^{-1}g^{\top}(x)\left((J_d(x)-R_d(x))\frac{\partial H_d}{\partial x}(x)-(J(x)-R(x))\frac{\partial H}{\partial x}(x)\right)}_{\beta(x)}+v
\end{equation}
locally stabilizes system \eqref{eq:phmechanical} and the closed-loop system \eqref{eqn:system2} is passive with the new input $v$ and output $y$. Moreover, if $x^*$ is the largest invariant subset in $\{x\in\mathcal{X}|\frac{\partial H_d^\top}{\partial x}(x)R_d(x)\frac{\partial H_d}{\partial x}(x)=0\}$ then $x^*$ is  asymptotically stable. The results holds globally  if $x^*$ is a global minimum of $H_d(x)$ and $H_d(x)$ radially unbounded.
\end{theorem} 
After some straightforward algebraic manipulations, it is easy to show that the desired functions can be rewritten as the sum of the functions  $J(x), R(x), H(x)$ of the pH system in \eqref{eq:phmechanical} and some \emph{auxiliary} functions $J_a(x), R_a(x), H_a(x)$ such that
\begin{equation}
    J_d(x):=J(x)+J_a(x),\quad  R_d(x):=R(x)+R_a(x) \quad H_d(x):=H(x)+H_a(x).
\end{equation}
These auxiliary terms are used to \emph{assign} the desired functions in  Theorem \ref{theorem:IDA-PBC}. Some implicit design requirements in Theorem \ref{theorem:IDA-PBC} are mentioned below as these are key for the constructive procedure that will be presented in Section 3.
\begin{itemize}
	\item[(P1)] \underline{Structure preservation}: the matrices  $[J(x) + J_a(x)]$ and $[R(x) + R_a(x)] \geq 0$ are  skew-symmetric and symmetric, respectively.
	\item[(P2)] \underline{Integrability}:  $K(x)=\frac{\partial H_a}{\partial x}(x)$ is the gradient of a scalar function, i.e. $\frac{\partial K}{\partial x}(x) = \left[\frac{\partial K}{\partial x}(x)\right]^\top.$
	\item[(P3)] \underline{Equilibrium assignment}: $\frac{\partial H_d}{\partial x}(x^\star) = \frac{\partial H}{\partial x}(x^\star) + \frac{\partial H_a}{\partial x}(x^\star) = 0$, where $x^\star = \textrm{argmin}_{x\in\Omega}H_d(x).$
	\item[(P4)] \underline{Lyapunov stability} $H_d(x)$ is a positive definite function at $x^\star$, e.g. $\frac{\partial^2 H}{\partial x^2}(x^\star) + \frac{\partial^2 H_a}{\partial x^2}(x^\star) > 0.$
\end{itemize}
Under these conditions, the closed-loop system will be a PCH system with the form \eqref{eqn:system2}, where $H_d(x)$ is the closed-loop energy function.

\section{Neural IDA-PBC}\label{sec:NeuralIDA}
In this section we introduce how Neural Networks can be used to solve the PDEs in \eqref{eqn:matchingeqs}. This methodology is based on the intrinsic property of Neural Networks as Universal Approximators \citep{hornik_multilayer_1989,cybenko_approximation_1989, kratsios_universal_2021}.

\subsection{Using PINNs to solve PDEs}\label{sec:PINN}
As showcased in \citep{raissi_physics_2017, raissi_physics_2017-1}, the universal approximation property of NNs can be exploited to find the solution of ill-posed problems such as the ones involving PDEs. Considering a nonlinear parametric partial differential equation of the form 
\begin{align}
	u_t + \mathcal{N}[u;\lambda] = 0,
	\label{eqn:nonlinearPDE}
\end{align}
where $\mathcal{N}[\cdot; \lambda]$ is a nonlinear operator parameterized by $\lambda$ and $u_t$ is the time derivative of the function $u(t,x)$. We can define a residual term, $f(t,x)$, as the left-hand-side of \eqref{eqn:nonlinearPDE} such that the NN that minimizes this residual is considered an approximated solution of \eqref{eqn:nonlinearPDE}, i.e. 
\begin{align}
	\begin{split}
		\mathcal{L}(\theta; x) :&= u_t(\theta;t,x) + \mathcal{N}[u(\theta;t,x);\lambda], \qquad 
		\theta^\star = \mathrm{argmin}_\theta \mathcal{L}(\theta; x).
	\end{split}\label{eqn:PINNminimize}
\end{align}
In this case, $u(t, x) = u_\theta(\theta^\star; t, x) + \epsilon$, and the result is a deep neural network that encodes the relation(s) of PDE(s). This methodology is known as \textit{Physics Informed Neural Networks (PINNs)}.

\subsection{Using PINNs to solve the non-parameterized IDA-PBC}
Assuming the existence of solutions for the problem defined by \eqref{eqn:matchingeqs}, (P1) - (P4), finding the missing functions $J_a(x)$, $R_a(x)$ and $H_a(x)$ is an inverse problem, in particular an ill-posed one, which makes it a perfect candidate to be studied under the scope of physics-informed machine learning \citep{karniadakis_physics-informed_2021}.

\begin{figure}[h]
    \centering
    \includegraphics[width=8.5cm]{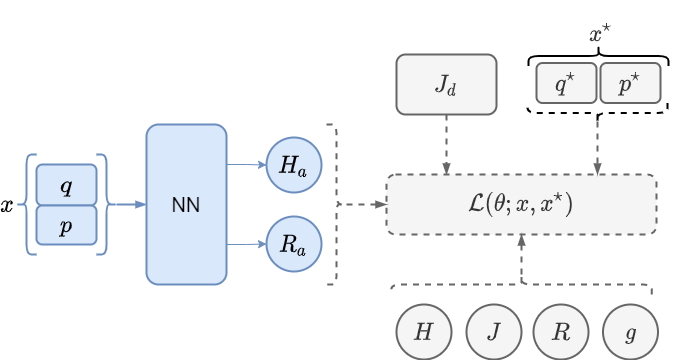}
    \caption{Neural IDA-PBC diagram. The NN is used to solve the non-parameterized IDA-PBC problem. It approximates the auxiliary energy and damping functions, $H_a$, and $R_a$, such that the loss function $\mathcal{L}(\theta; x, x^\star)$ is minimized. This loss function is methodologically constructed in Section \ref{sec:IDAresiduals}. $x$ is a tuple of points in the state space $\mathcal{X}$. $H$, $J$, $R$ and $g$ are the open loop Hamiltonian, interconnection, damping and input matrices as in Section \ref{sec:Preliminaries-pH}. The desired interconnection matrix $J_d(x)$ and desired equilibrium point $x^\star$ are fixed and passed directly into the loss function.}
    \label{fig:NNstruct}
\end{figure}

To reduce the underdeterminism of the problem, we adopt the non-parameterized IDA-PBC approach, formalized in \citep{nageshrao_adaptive_2015}, which requires solving the MEs after fixing the choice of $J_d(x)$ and $R_d(x)$. However, in order to avoid triviality when choosing these values, we propose to only fix $J_d(x)$, while iteratively adapting $R_d(x)$ to obtain certain desired properties of the closed-loop system. At the same time, $H_d(x):= H(x) + H_a(x)$ can be approximated using NNs. The diagram in Figure \ref{fig:NNstruct} illustrates this working strategy. 

\subsection{IDA-PBC loss function}\label{sec:IDAresiduals}
This section defines the residuals that encode the IDA-PBC principles listed in Section \ref{sec:IDA}. They are then combined into the objective function needed to be minimized using NNs. For simplicity in the presentation, we will only consider fully-actuated mechanical systems given by \eqref{eq:phmechanical} with the standard Hamiltonian function $H$. The residuals are created from (P1) - (P4) as follows.

\paragraph{(P1) \underline{Structure preservation residual}} 
\subparagraph{Interconnection matrix} In order to satisfy this property we are forced to choose an auxiliary interconnection matrix $J_a(x)$ with a skew-symmetric form. Correspondingly, we consider the following structure for $J_a$ as proposed in \citep{ortega_stabilization_2000},	$J_a(x) = \sbm{0 & J_1(x)\\ -J_1(x)^\top & J_2(x)},$ where $J_1(x)$ is an arbitrary $n\times n$ matrix and $J_2(x)\in\mathbb{R}^{n\times n}$ is skew-symmetric matrix. Both are fixed by the user prior to the optimization process.

\subparagraph{Dissipation matrix} The desired damping matrix $R_d$ in (P1) must be positive semi-definite. For this matrix, we also choose to adopt the parametric version of the structure in \citep{ortega_stabilization_2000}, $R_a(\theta;x) = \sbm{0 & 0 \\0 & R_2(\theta;x)}$, where $R_2(\theta;x)\in\mathbb{R}^{n\times n}$ is symmetric and positive-definite matrix parametrized by $\theta$. The parameterization of $R_a$ is to enable the discovery of admissible damping functions that fulfill the design criteria during the optimization process. For instance, the chosen criteria is related to the transient behavior of the closed-loop system, where we can prescribe the transient behaviour through the enforcement of the rate of energy decrease via $F_d(x) = J_d(x) - R_d(x)$. In particular, we define the following residual
\begin{equation}
    f_\mathrm{transient}(\theta;x) := \underbrace{\max\left\{0, c+\mathrm{Re}(\sigma(F_d(x))\right\}}_{f_\mathrm{damping}} + \underbrace{\|\mathrm{Im}(\sigma(F_d(x))\|}_{f_\mathrm{harmonic}},
	\label{eqn:residualtransient}
\end{equation}
where $c>0$ is the prescribed convergence rate of the new energy function $H_d$, Re and Im are the real and imaginary part, respectively. The first term is related to the residual for prescribing the damping of target system while the second term corresponds to the residual for minimizing the harmonic oscillation.

\paragraph{(P2) \underline{Integrability residual}} 
As the NNs are used to obtain the scalar function $H_a(\theta;x)$, the integrability condition (P2) is fulfilled by construction. 

\paragraph{ (P3) \underline{Equilibrium assignment residual}} 
The condition in (P3) implies that the closed-loop energy function, has a global minimum at the desired equilibrium $x^\star = (q^\star,0)$. Correspondingly, we consider the following residual function to assign the equilibrium point
\begin{align}
	f_\mathrm{eq}(\theta;x,x^*) := \underbrace{\left\|\frac{\partial H_d}{\partial x} (\theta; x^\star)\right\|^2}_{f_{\mathrm{eq},1}} +\underbrace{(H_d (\theta;x^\star))^2 + \max\{0, -H_d(\theta;x)\}}_{f_{\mathrm{eq},2}},
	\label{eqn:residualequilibrium}
\end{align} 
where $f_{\mathrm{eq},1}$ defines that $x^*$ is a stationary point and $f_{\mathrm{eq},2}$ is used to induce a lower bound on $H_d$. 

\paragraph{(P4) \underline{Lyapunov stability residual}} 
This condition is equivalent to having the spectrum of $\frac{\partial^2 H_d}{\partial x^2}(\theta;x)$ to be positive. Therefore the residual function to guarantee (P4) is given by 
\begin{align}
	f_\mathrm{lyap} := \max\left\{0, c-\sigma(H_d(\theta;x))\right\},
	\label{eqn:residuallyapunov}
\end{align} 
where $\sigma$ denotes the spectrum of a matrix and $c>0$ is again a constant value that can be added to make this condition stricter.

\paragraph{\underline{IDA-PBC matching equation residual }} 
We can rewrite the IDA-PBC ME \eqref{eqn:matchingeqs} in the residual form moving all the terms to the left-hand-side as it is presented in  \ref{sec:PINN},
\begin{align}
	f_{\mathrm{matching}} := g^\perp(x)[J_d(x)-R_d(x)]\frac{\partial H_d}{\partial x}(x) - g^\perp(x)[J(x)-R(x)]\frac{\partial H}{\partial x}(x). 	\label{eqn:residualmatching}
\end{align}\vspace{0.2cm}

Finally, we can combine all residuals \eqref{eqn:residualtransient} - \eqref{eqn:residualmatching} to get the final cost function
\begin{align}
	\mathcal{L} = f_{\mathrm{transient}} + f_\mathrm{eq} + f_\mathrm{lyap} + f_\mathrm{matching}. 
	\label{eqn:IDAresidualfunction}
\end{align} 

This cost function implicitly encodes the family of admissible solutions of the IDA-PBC control scheme. As a consequence of the universal approximation theorem \citep{hornik_multilayer_1989, cybenko_approximation_1989}, if follows that the NN that minimizes \eqref{eqn:IDAresidualfunction} approximates the required auxiliary functions  $R_a^\star(x) \approx R_a(\theta^\star;x) + \epsilon$ and $H_a^\star(x) \approx H_a(\theta^\star;x) + \epsilon$, that asymptotically stabilize the system at the desired equilibrium point $x^\star$.

\section{Neural IDA-PBC for mechanical systems}\label{sec:Simulations}
This section includes the numerical results for two different examples. The NN used throughout this section has 3 hidden layers of 20 units each, and the optimizer chosen is Adam with a learning rate of 0.001, that is interrupted after certain convergence tolerance is reached, followed by L-BFGS-B as presented in the seminal work \citep{raissi_physics_2017}.

\subsection{Simple Pendulum}
The simple pendulum is a well established example in robotics due to its non-linear dynamics and simple formulation. In this case we consider a system with no natural damping, actuated at its joint i.e. $g(x) = \begin{bmatrix}
	0 & 1
\end{bmatrix}^\top$, to be stabilized at an arbitrary configuration $x^\star = (q^\star, 0)$. The total energy of this system is given by $H(x) = \frac{1}{2ml^2}p^2 + mgl(1-\cos{q})$. 

\begin{figure}[ht]
\centering
\subfigure[Potential maps of different energy functions for $J_1=\textrm{diag}(-0.5)$.]{{%
\includegraphics[width=.4\textwidth]{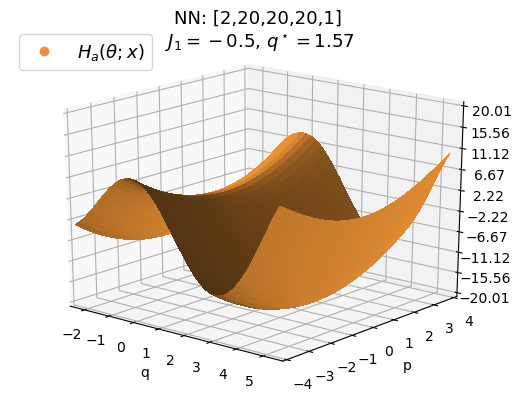}
\includegraphics[width=.4\textwidth]{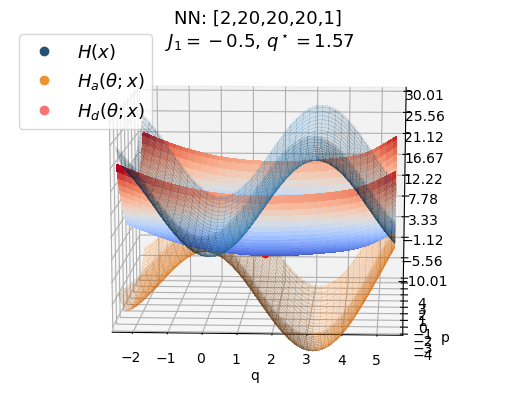}
\label{fig:SP-Hs-Jd05}}}
\subfigure[Temporal dynamics of $q$ (left) and $p$ (right) for different initial conditions.]{{%
\includegraphics[width=.4\textwidth]{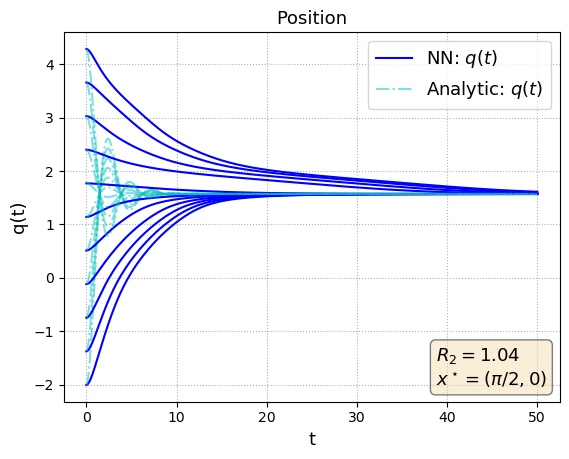}
\includegraphics[width=.4\textwidth]{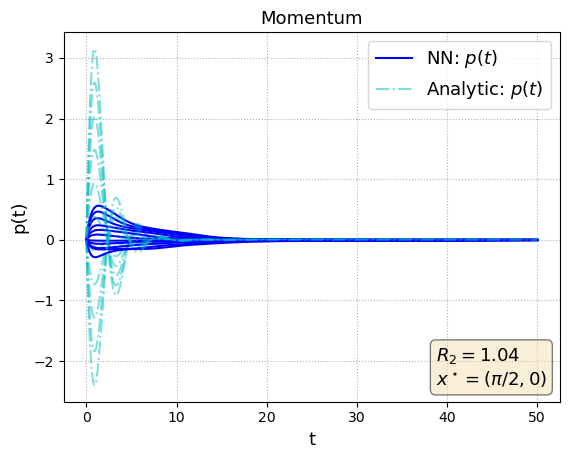}
\label{fig:SP-T-Jd05}}}
\caption{Numerical simulations of the simple pendulum system for different values of $J_a(x)$ parameterized as in (P1): (a) On the left, the approximated $H_a(\theta;x)$ that satisfies the simple pendulum matching equations when $J_1=\textrm{diag}(-0.5)$. On the right, the resulting closed-loop energy function $H_d(\theta;x)=H(x)+H_a(\theta;x)$ (solid curve) which is convex and has a minimum at $x^\star = (\pi/2,0)$; (b) Neural IDA-PBC (solid line) after fixing $J_1=-0.5$ vs the analytical solution of the MEs (dashed line) when $J_a(x)=0$ and the potential energy is compensated to be quadratic around $x^\star$.}
\label{fig:SP-Jd05}
\end{figure}

\begin{figure}[ht]
\centering
\subfigure[Potential maps of different energy functions for $J_1=\textrm{diag}(1.0)$.]{%
\includegraphics[width=.4\textwidth]{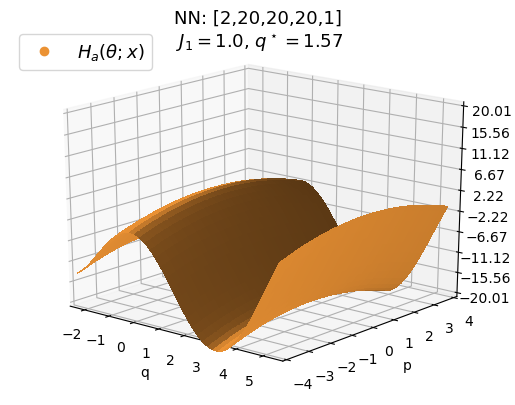}
\hfill
\includegraphics[width=.4\textwidth]{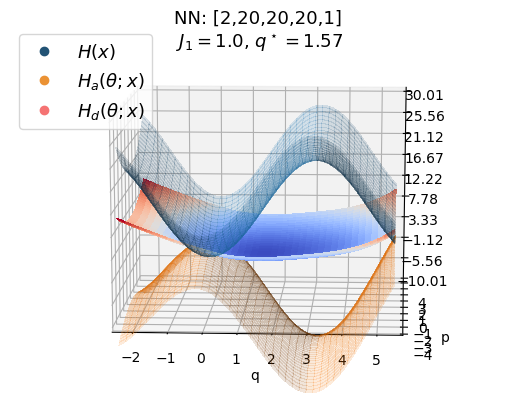}
\label{fig:SP-Hs-Jd2}}
\subfigure[Temporal dynamics of $q$ (left) and $p$ (right) for different initial conditions.]{%
\includegraphics[width=.4\textwidth]{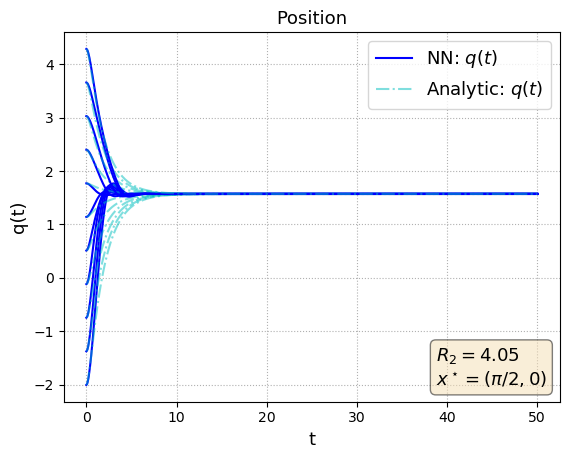}
\hfill
\includegraphics[width=.4\textwidth]{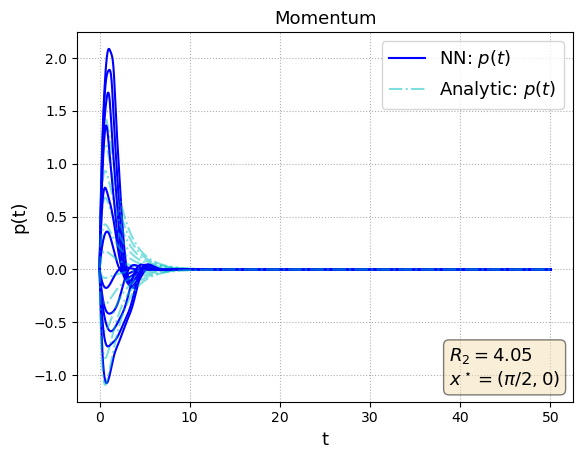}
\label{fig:SP-T-Jd2}}
\caption{Numerical simulations of the simple pendulum system for different values of $J_a(x)$ parameterized as in (P1): (a) On the left, the approximated $H_a(\theta;x)$ that satisfies the simple pendulum matching equations when $J_1=\textrm{diag}(1.0)$. On the right, the resulting closed-loop energy function $H_d(\theta;x)=H(x)+H_a(\theta;x)$ (solid curve) which is convex and has a minimum at $x^\star = (\pi/2,0)$; (b) Neural IDA-PBC (solid line) after fixing $J_1=1.0$ vs the analytical solution of the MEs (dashed line) when $J_a(x)=0$ and the potential energy is compensated to be quadratic around $x^\star$.}
\label{fig:SP-Jd2}
\end{figure}

Substituting into \eqref{eqn:matchingeqs}, we obtain the ME for this problem $(1+J_1)\frac{\partial H_a}{\partial p}(x) = -J_1\frac{\partial H}{\partial p}$, which together with (P1) - (P4) encode the family of solutions defined by the IDA-PBC control methodology. In Figures \ref{fig:SP-Jd05} and \ref{fig:SP-Jd2} we present the numerical results for two illustrative cases where $J_1=-0.5$ and $J_1= 1$ respectively. The choice of these values can be intuitively interpreted changing the closed-loop inertia by a factor of two. 

In both cases we observe that the NN is able to approximate the required $H_a(x)$ and $R_a(x)$, therefore asymptotically stabilizing the closed-loop system at the desired point $x^\star = (\pi/2,0)$. Additionally, Figures \ref{fig:SP-Jd05}(b) and \ref{fig:SP-Jd2}(b) show a comparison of $q(t)$ and $p(t)$ between Neural IDA-PBC methodology for a non-trivial $J_a(x)$ and the solution of IDA-PBC when $J_1=0$ is chosen to simplify the original MEs (pure potential energy compensation). From this comparison, it is evident that although (in the simple pendulum case) modifying the kinetic energy of the system is not crucial for stabilization, being able to solve the MEs for non-arbitrary choices of $Ja(x)$ is important to not limit the energy-based controller to a particular family of solutions where the trajectory might be sub-optimal.

\subsection{Double Pendulum}
The double pendulum system is also very common in robotics and it is normally used to represent a 2-link manipulator. In this example we also consider that there is no natural damping on the system, that the actuation also occurs at its joints, i.e. $g(x) = \begin{bmatrix}
	0_{2\times 2} & I_2
\end{bmatrix}^\top$ and we want to stabilize the system at $x^\star = ({q_1}^\star,{q_2}^\star, 0,0)$. The total energy function for this system is given by $H(q,p)=\frac{1}{2}\sbm{p_1&p_2}M^{-1}(q)\sbm{p_1\\ p_2}+(m_1+m_2)gl_1(1-\cos(q_1)) + m_2gl_2(1-\cos(q_2))$ where the inertia matrix $M(q)$ is given by $M(q)=\sbm{ (m_1+m_2)l_1^2 & m_2l_1l_2\cos(q_1-q_2)\\ m_2l_1l_2\cos(q_1-q_2) & m_2l_2^2}$. 

In this case, substituting into \eqref{eqn:matchingeqs}, we obtain two MEs that need to be satisfied in order to obtain an IDA-PBC control law that stabilizes the system: $$(1+J_{11})\frac{\partial H_a}{\partial p_1} H_a(x) = -J_{11}\frac{\partial H_a}{\partial p_1}(x) \quad\text{  and  }\quad(1+J_{12})\frac{\partial H_a}{\partial p_2}(x) = -J_{12}\frac{\partial H_a}{\partial p_2}(x).$$
Notice that these MEs, only impose a structure in the generalized momenta directions. Therefore, in order to ease the discovery of the solution using NNs, we create an additional complementary residual to steer the closed-loop system in the generalized coordinates direction. Suppose $H_d(x) = \frac{1}{2}p^\top N(q)p + U_d(q-q^\star)$, where $N(q)$ is a positive definite matrix, then the gradient in the coordinates direction $\frac{\partial H_d}{\partial q} (x) = F(q,p) +   \frac{\partial U_d}{\partial q}(q-q^\star) \approx G_d(q-q^\star)$ must be a monotonic function to guarantee the Lyapunov stability condition (P4). We can choose this function apriori, by  defining a complementary residual as
\begin{align*}
    f_\mathrm{comp}(\theta;x,x^*) :&= G_d(q-q^\star) - \left(\frac{\partial H}{\partial q}(x) +\frac{\partial H_a}{\partial q}(\theta;x) \right)\\
	\mathcal{L} &= f_{\mathrm{transient}} + f_\mathrm{eq} + f_\mathrm{lyap} + f_\mathrm{matching} + \lambda f_\mathrm{comp}
\end{align*}
where $\lambda$ is an extra multiplier that can be adjusted to change the closed-loop temporal dynamics.

\begin{figure}[ht]
\centering
\subfigure[The temporal dynamics of $q$ (left) and $p$ (right) for different initial conditions.]{%
\includegraphics[width=.4\textwidth]{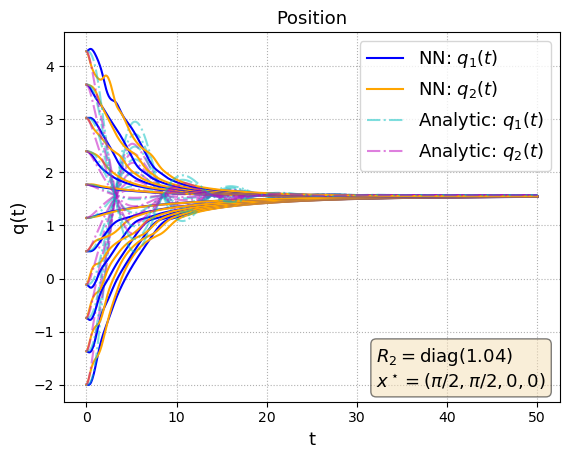}
\hfill
\includegraphics[width=.4\textwidth]{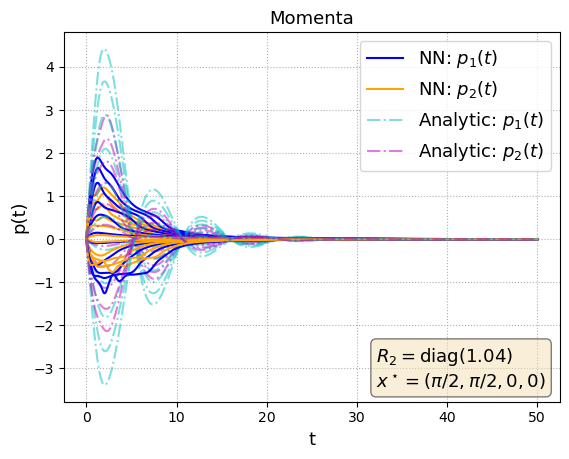}
\label{fig:DP-T-Jd05}}
\subfigure[The temporal dynamics of $q$ (left) and $p$ (right) for different initial conditions.]{%
\includegraphics[width=.4\textwidth]{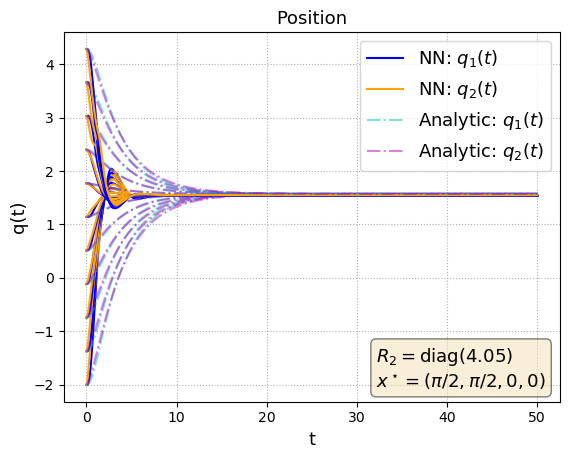}
\hfill
\includegraphics[width=.4\textwidth]{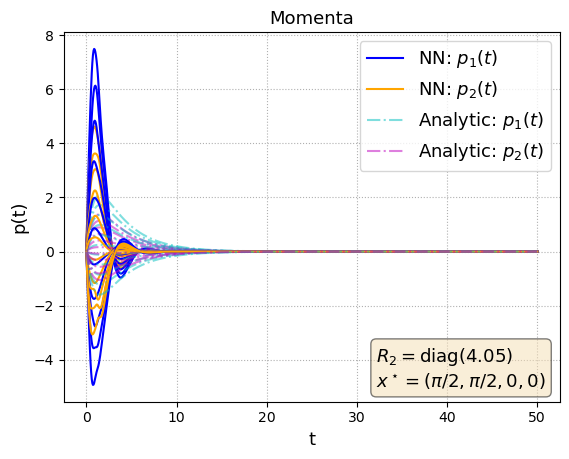}
\label{fig:DP-T-Jd2}}
\caption{Numerical simulations of the double pendulum system for different values of $J_a(x)$ parameterized as in (P1): (a) Neural IDA-PBC (solid line) after fixing the pair  ($J_1=\textrm{diag}(-0.5)$, $J_2 = 0$) vs the analytical solution of the MEs (dashed line) when $J_a(x) = 0$ and the potential energy is compensated to be quadratic around $x^\star$; (b) Neural IDA-PBC (solid line) after fixing the pair  ($J_1=\textrm{diag}(1.0)$, $J_2 = 0$) vs the analytical solution of the MEs (dashed line) when $J_a(x) = 0$ and the potential energy is compensated to be quadratic around $x^\star$.}
\label{fig:DP-T}
\end{figure}

In Figure \ref{fig:DP-T} we present the temporal dynamics of $q(t)$ and $p(t)$ for two different cases $(J_{1} = \textrm{diag}(-0.5), J_2 = 0)$ and $(J_{1} = \textrm{diag}(1.0), J_2 = 0)$. Here, we also compare the Neural IDA-PBC response with the case where $J_a(x) = 0$ (pure potential energy compensation). 
We can see that the Neural IDA-PBC method is capable of asymptotically stabilizing the system around a desired point. And again, the comparison with the pure potential energy compensation case (as it is commonly done in practice), shows how modifying the closed-loop interconnection matrix has a big impact in the closed-loop system trajectories.

\section{Conclusions}\label{sec:Conclusions}

In this paper we have presented a systematic approach to solve the IDA-PBC problem formulation using control-informed neural networks. For this purpose we have adapted the residual methodology presented in \citep{karniadakis_physics-informed_2021} to cover the fundamentals required by the IDA control method and the PBC theory. Despite the fact that fully-actuated systems do not require  kinetic energy-shaping for stabilization purposes,
the simulation results show that modifying the kinetic energy is beneficial to improve the transient behavior. However, finding the controller that achieves this and satisfies the MEs can be very hard  due to the nonlinearities and coupled dynamics of the system under study. The main advantage of Neural IDA-PBC is that it enables finding approximations to this type of controllers thanks to the automatic discovery of solutions to the MEs. 

Conclusively, the presented methodology enables the derivation of \emph{total energy shaping controllers} that allow exploiting the true physics of the system without the need of taking \emph{restrictive} mathematical assumptions to ease the calculations. 

\acks{This publication is part of the project Digital Twin project 6 with project number P18-03 of the research programme Perspectief which is (mainly) financed by the Dutch Research Council (NWO).}

\bibliography{L4DCNEURALIDAPBC}

\begin{thebibliography}{20}
\providecommand{\natexlab}[1]{#1}
\providecommand{\url}[1]{\texttt{#1}}
\expandafter\ifx\csname urlstyle\endcsname\relax
  \providecommand{\doi}[1]{doi: #1}\else
  \providecommand{\doi}{doi: \begingroup \urlstyle{rm}\Url}\fi

\bibitem[Chang and Gao(2021)]{chang_stabilizing_2021}
Ya-Chien Chang and Sicun Gao.
\newblock Stabilizing {Neural} {Control} {Using} {Self}-{Learned} {Almost}
  {Lyapunov} {Critics}.
\newblock \emph{arXiv:2107.04989}, July 2021.

\bibitem[Chang et~al.(2020)Chang, Roohi, and Gao]{chang_neural_2020}
Ya-Chien Chang, Nima Roohi, and Sicun Gao.
\newblock Neural {Lyapunov} {Control}.
\newblock \emph{arXiv:2005.00611}, December 2020.

\bibitem[Cybenko(1989)]{cybenko_approximation_1989}
G.~Cybenko.
\newblock Approximation by superpositions of a sigmoidal function.
\newblock \emph{Math. Control Signal Systems}, 2\penalty0 (4):\penalty0
  303--314, December 1989.

\bibitem[Fujimoto et~al.(2003)Fujimoto, Horiuchi, and
  Sugie]{fujimoto_optimal_2003}
K.~Fujimoto, T.~Horiuchi, and T.~Sugie.
\newblock Optimal control of {Hamiltonian} systems with input constraints via
  iterative learning.
\newblock In \emph{42nd {IEEE} {International} {Conference} on {Decision} and
  {Control}}, pages 4387--4392, 2003.

\bibitem[Grüne(2020)]{grune_computing_2020}
Lars Grüne.
\newblock Computing {Lyapunov} functions using deep neural networks.
\newblock \emph{arXiv:2005.08965}, November 2020.

\bibitem[Hornik et~al.(1989)Hornik, Stinchcombe, and
  White]{hornik_multilayer_1989}
Kurt Hornik, Maxwell Stinchcombe, and Halbert White.
\newblock Multilayer feedforward networks are universal approximators.
\newblock \emph{Neural Networks}, 2\penalty0 (5):\penalty0 359--366, January
  1989.

\bibitem[Isidori et~al.(1995)Isidori, Sontag, and Thoma]{isidori1995nonlinear}
Alberto Isidori, ED~Sontag, and M~Thoma.
\newblock \emph{Nonlinear control systems}, volume~3.
\newblock Springer, 1995.

\bibitem[Karniadakis et~al.(2021)Karniadakis, Kevrekidis, Lu, Perdikaris, Wang,
  and Yang]{karniadakis_physics-informed_2021}
George~Em Karniadakis, Ioannis~G. Kevrekidis, Lu~Lu, Paris Perdikaris, Sifan
  Wang, and Liu Yang.
\newblock Physics-informed machine learning.
\newblock \emph{Nat Rev Phys}, 3\penalty0 (6):\penalty0 422--440, June 2021.

\bibitem[Khalil(2015)]{khalil_nonlinear_2015}
Hassan~K. Khalil.
\newblock \emph{Nonlinear control}.
\newblock Pearson, Boston, 2015.

\bibitem[Kratsios and Papon(2021)]{kratsios_universal_2021}
Anastasis Kratsios and Leonie Papon.
\newblock Universal {Approximation} {Theorems} for {Differentiable} {Geometric}
  {Deep} {Learning}.
\newblock \emph{arXiv:2101.05390}, June 2021.

\bibitem[Massaroli et~al.(2021)Massaroli, Poli, Califano, Park, Yamashita, and
  Asama]{massaroli_optimal_2021}
Stefano Massaroli, Michael Poli, Federico Califano, Jinkyoo Park, Atsushi
  Yamashita, and Hajime Asama.
\newblock Optimal {Energy} {Shaping} via {Neural} {Approximators}.
\newblock \emph{arXiv:2101.05537}, January 2021.

\bibitem[Nageshrao et~al.(2015)Nageshrao, Lopes, Jeltsema, and
  Babusˇka]{nageshrao_adaptive_2015}
S~P Nageshrao, G~A~D Lopes, D~Jeltsema, and R~Babusˇka.
\newblock Adaptive and {Learning} {Control} of port-{Hamiltonian} {Systems}:
  {A} {Survey}.
\newblock \emph{IEEE Transactions on Automatic Control}, page~37, 2015.

\bibitem[Ortega and Spong(2000)]{ortega_stabilization_2000}
Romeo Ortega and Mark~W. Spong.
\newblock Stabilization of {Underactuated} {Mechanical} {Systems} {Via}
  {Interconnection} and {Damping} {Assignment}.
\newblock \emph{IFAC Proceedings Volumes}, 33\penalty0 (2):\penalty0 69--74,
  March 2000.

\bibitem[Ortega et~al.(2002)Ortega, van~der Schaft, Maschke, and
  Escobar]{ortega_interconnection_2002}
Romeo Ortega, Arjan van~der Schaft, Bernhard Maschke, and Gerardo Escobar.
\newblock Interconnection and damping assignment passivity-based control of
  port-controlled {Hamiltonian} systems.
\newblock \emph{Automatica}, 38\penalty0 (4):\penalty0 585--596, April 2002.

\bibitem[Raissi et~al.(2017{\natexlab{a}})Raissi, Perdikaris, and
  Karniadakis]{raissi_physics_2017}
Maziar Raissi, Paris Perdikaris, and George~Em Karniadakis.
\newblock Physics {Informed} {Deep} {Learning} ({Part} {I}): {Data}-driven
  {Solutions} of {Nonlinear} {Partial} {Differential} {Equations}.
\newblock \emph{arXiv:1711.10561}, November 2017{\natexlab{a}}.

\bibitem[Raissi et~al.(2017{\natexlab{b}})Raissi, Perdikaris, and
  Karniadakis]{raissi_physics_2017-1}
Maziar Raissi, Paris Perdikaris, and George~Em Karniadakis.
\newblock Physics {Informed} {Deep} {Learning} ({Part} {II}): {Data}-driven
  {Discovery} of {Nonlinear} {Partial} {Differential} {Equations}.
\newblock \emph{arXiv:1711.10566}, November 2017{\natexlab{b}}.

\bibitem[Richards et~al.(2018)Richards, Berkenkamp, and
  Krause]{richards_lyapunov_2018}
Spencer~M. Richards, Felix Berkenkamp, and Andreas Krause.
\newblock The {Lyapunov} {Neural} {Network}: {Adaptive} {Stability}
  {Certification} for {Safe} {Learning} of {Dynamical} {Systems}.
\newblock \emph{arXiv:1808.00924}, October 2018.

\bibitem[van~der Schaft(2007)]{sanz-sole_port-hamiltonian_2007}
Arjan van~der Schaft.
\newblock Port-{Hamiltonian} systems: an introductory survey.
\newblock In \emph{Proceedings of the {International} {Congress} of
  {Mathematicians}}, pages 1339--1365. European Mathematical Society Publishing
  House, May 2007.

\bibitem[van~der Schaft(2017)]{arjanl2}
Arjan van~der Schaft.
\newblock \emph{L2-Gain and Passivity Techniques in Nonlinear Control}.
\newblock Springer International Publishing, 2017.

\bibitem[Vu and Lef{\`e}vre(2018)]{vu_connection_2018}
N.~M.~Trang Vu and L.~Lef{\`e}vre.
\newblock A connection between optimal control and {IDA}-{PBC} design.
\newblock \emph{IFAC-PapersOnLine}, 51\penalty0 (3):\penalty0 205--210, January
  2018.

\end{thebibliography}

\end{document}